\documentclass{sig-alt-full}

\usepackage{amssymb}
\usepackage{amsfonts}
\usepackage{amsmath}
\usepackage{latexsym}
\usepackage{times}
\usepackage{color}

\usepackage{afterpage}

\newcommand{\comment}[1]{}

\newcommand{\ket}[1]{|#1  \rangle}
\newcommand{\bra}[1]{\langle#1 |}
\newcommand{\proj}[1]{|{#1} \rangle\langle {#1} |}

\newcommand{\set}[1]{\left\{ {#1} \right\}}
\newcommand{\paren}[1]{\left( {#1} \right)}
\newcommand{\floor}[1]{\left\lfloor{#1} \right\rfloor}

\newcommand{\supp}{\mathrm{supp}}

\newcommand{\myspan}{\mathrm{span}}

\newcommand{\ith}{i^{th}}
\newcommand{\jth}{j^{th}}

\newcommand{\tr}{\mbox{Tr}}

\newcommand{\ze}{\mathbb{Z}}

\newcommand{\mypar}[1]{\vspace{6pt} \noindent {\bf #1}.\ }

\newtheorem{theorem}{Theorem}[section]
\newtheorem{lemma}[theorem]{Lemma}
\newtheorem{prop}[theorem]{Proposition}

\newtheorem{protocol}{Protocol}
\newtheorem{algm}{Algorithm}
 
\newtheorem{definition}{Definition}

\newcommand{\thmref}[1]{Theorem~\ref{thm:#1}}
\newcommand{\lemref}[1]{Lemma~\ref{lem:#1}}

\newcommand{\propref}[1]{Proposition~\ref{prop:#1}}
\newcommand{\remref}[1]{Remark~\ref{rem:#1}}

\newcommand{\secref}[1]{Section~\ref{sec:#1}}

\newcommand{\stepref}[1]{Step~\ref{step:#1}}
\newcommand{\protref}[1]{Protocol~\ref{prot:#1}}
\newcommand{\algref}[1]{Algorithm~\ref{alg:#1}}

\newcommand{\tgood}{\mbox{2{\sc -good}}}
\newcommand{\vqss}{{\sc vqss}}
\newcommand{\css}{{\sc css}}
\newcommand{\cvss}{{\sc vss}}
\newcommand{\ftqc}{{\sc ftqc}}
\newcommand{\mpqc}{{\sc mpqc}}
\newcommand{\ttp}{{\cal TTP}}
\newcommand{\A}{{\cal A}}
\newcommand{\D}{\mathcal{D}}
\newcommand{\E}{{\cal E}}
\newcommand{\C}{{\cal C}}

\newcommand{\TC}{{\tilde {{\cal C}}}}

\newcommand{\Sim}{{\cal S}}

\newcommand{\ri}{\mathcal{R}^I}

\newcommand{\fn}{{\cal F}^{\otimes n}} 
\newcommand{\F}{{\cal F}}

\newcommand{\vq}{V^{(q)}} \newcommand{\vp}{V^\perp}
 \newcommand{\vbq}{V_B^{(q)}}

\newcommand{\wq}{W^{(q)}} 

 \newcommand{\wbq}{W_B^{(q)}}

\def\compactify{\leftmargin=9pt \itemsep=0pt \topsep=0pt \partopsep=0pt \parsep=0pt}
\let\latexusecounter=\usecounter
\newenvironment{naritem}
  {\def\usecounter{\compactify\latexusecounter}
   \begin{itemize}}
  {\end{itemize}\let\usecounter=\latexusecounter}
\newenvironment{narenum}
  {\def\usecounter{\compactify\latexusecounter}
   \begin{enumerate}}
  {\end{enumerate}\let\usecounter=\latexusecounter}
\newenvironment{nardesc}
  {\def\usecounter{\compactify\latexusecounter}
   \begin{description}}
  {\end{description}\let\usecounter=\latexusecounter}

\def\compactifyB{\leftmargin=12pt \itemsep=0pt \topsep=0pt \partopsep=0pt \parsep=0pt}
\let\latexusecounter=\usecounter
\newenvironment{naritem*}
  {\def\usecounter{\compactifyB\latexusecounter}
   \begin{itemize}}
  {\end{itemize}\let\usecounter=\latexusecounter}
\newenvironment{narenum*}
  {\def\usecounter{\compactifyB\latexusecounter}
   \begin{enumerate}}
  {\end{enumerate}\let\usecounter=\latexusecounter}
\newenvironment{nardesc*}
  {\def\usecounter{\compactifyB\latexusecounter}
   \begin{description}}
  {\end{description}\let\usecounter=\latexusecounter}

\renewenvironment{proof}{\noindent \textbf{Proof}:}{$\Box$ \vskip 12pt}

\newcommand{\an}[1]{}
\newcommand{\dn}[1]{}
\newcommand{\cn}[1]{}

\begin{document}

\conferenceinfo{STOC'02,} {May 19-21, 2002, Montreal, Quebec, Canada.}

\CopyrightYear{2002}

\crdata{1-58113-495-9/02/0005}

\title{Secure Multi-party Quantum Computation}


\numberofauthors{3}
%

\author{
%
  \alignauthor Claude Cr{\'e}peau\titlenote{Supported 
by Qu\'ebec's FCAR and Canada's NSERC.}\\
  \affaddr{McGill University}\\
  \email{{\normalsize \sf crepeau@cs.mcgill.ca}}
  \alignauthor Daniel Gottesman\titlenote{Supported by the Clay Mathematics Institute.}\\
  \affaddr{UC Berkeley}\\
  \email{{\normalsize \sf gottesma@eecs.berkeley.edu}}
  \alignauthor Adam Smith\titlenote{Supported by the US DoD MURI program administered by the Army Research Office under grant DAAD19-00-1-0177.}\\
  \affaddr{MIT}\\
  \email{{\normalsize \sf asmith@theory.lcs.mit.edu}} 
}

\date{}
\maketitle

[This version appears with the permission of the ACM. Only minor typographical changes have been made from the version which appeared in the proceedings of {\em STOC 2002}.]

\begin{abstract}  
  \emph{Secure multi-party computing}, also called \emph{secure
   function evaluation}, has been extensively studied in classical
  cryptography. We consider the extension of this task to computation
  with quantum inputs and circuits. Our protocols are
  information-theoretically secure, i.e. no assumptions are made on
  the computational power of the adversary. For the 
  weaker task of \emph{verifiable quantum secret sharing}, we give a protocol
  which tolerates any $t < n/4$ cheating parties (out of $n$).  This
  is shown to be optimal. We use this new tool to show how to perform
  any multi-party quantum computation as long as the number of
  dishonest players is less than $n/6$.
\end{abstract}




\pagenumbering{arabic}

\section{Introduction}
\label{sec:intro}

Secure distributed protocols have been an important and fruitful area
of research for modern cryptography.  In this setting, there is a
group of participants who wish to perform some joint task, despite the
fact that some of the participants in the protocol may cheat in order
to obtain additional information or corrupt the outcome.

We investigate a quantum version of an extensively studied classical
problem, \emph{secure multi-party computation} (or \emph{secure
function evaluation}), first introduced by \cite{GMW87}.
%
%
A multi-party quantum computing (\mpqc) protocol allows $n$
participants $P_{1},P_{2},\ldots,P_{n}$ to compute an $n$-input
quantum circuit in such a way that each party $P_{i}$ is responsible
for providing one of the input states. The output of the
circuit is broken into $n$ components ${\cal
  H}_{1}\otimes\ldots\otimes{\cal H}_{n}$, and $P_{i}$ receives
the output ${\cal H}_{i}$.  
Note that the inputs to this protocol are arbitrary quantum
states---the player providing an input need only have it in his
possession; he does not need to know a classical description of it.
Moreover, unlike in the classical case, we cannot assume without loss
of generality that the result of the computation will be broadcast.
Instead, each player in the protocol receives some part of the output.

Informally, we require two security conditions:\\ 
\emph{- Soundness and Completeness:}
no coalition of $t$ or fewer cheaters should be able to affect
the outcome of the protocol beyond their ability to choose their
inputs.\\
\emph{- Privacy:}
no coalition of $t$ or fewer cheaters should learn anything beyond
what they can deduce from their initial knowledge of their input and
from their part of the output.

\mypar{Verifiable Quantum Secret Sharing}
In order to construct \mpqc\ protocols, we consider a subtask which we
call \emph{verifiable quantum secret sharing}. In classical
cryptography, a verifiable secret sharing scheme \cite{ChorGMA85} is a
two phase protocol with one player designated as the ``dealer''. After
the first phase (\emph{commitment}), the dealer shares a secret
amongst the players. In the second phase (\emph{recovery}), the
players reconstruct the value publicly.

The natural quantum version of this allows a dealer to share a state
$\rho$ (possibly unknown to him but nonetheless in his possession).
Because quantum information cannot be cloned, we cannot require that
the state be reconstructed publicly; instead, the recovery phase also
has a designated player, the reconstructor $R$. We require that,
despite any malicious actions by
$\leq t$ players:\\
\emph{- Soundness:} As long as $R$ is honest and the dealer passes
  the commitment phase successfully, then there is a unique quantum
  state which can be recovered by $R$.\\
\emph{- Completeness:} When $D$ is honest, then he always passes
  the commitment phase. Moreover, when $R$ is also honest, then the
  value recovered by $R$ is exactly $D$'s input $\rho$.\\
\emph{- Privacy:} When $D$ is honest, no other player learns info
  about $D$'s input until the recovery step.

Note that for quantum data, the privacy condition 
is redundant
: any information
obtained about the shared state would imply some
disturbance of that state,
contradicting the completeness requirement.

\mypar{Contributions}
We give a protocol for verifiable quantum secret sharing that
  tolerates any number $t<n/4$ of cheaters.
We show that this is optimal, by proving that \vqss\ is
  impossible when $t\geq n/4$.
Based on techniques from fault-tolerant quantum computing, we
  use our \vqss\ protocol to construct a multi-party quantum
  computation protocol tolerating any $t<n/6$ cheaters.
({\sc Mpqc} is similar to standard fault-tolerance but with a different 
error model,
see Previous Work).
%
Our protocols run in time polynomial in both $n$, the number of
players, and $k$, the security parameter. The error of the protocols
is exponentially small in $k$.

Beyond these specific results, there are a number of conceptual
contributions of this paper to the theory of quantum cryptographic
protocols.
We provide a simple, general framework for defining and proving
  the security of distributed quantum protocols in terms of
  equivalence to an ideal protocol involving a third party. This
  follows the definitions for classical multi-party protocols.
The analysis of our protocols leads us to consider various
  notions of local ``neighborhoods'' of quantum states, and more
  generally of quantum codes. We discuss three notions of a
  neighborhood. The notion most often used for the analysis of quantum
  error correction and fault-tolerance is insufficient for our needs,
  but we show that a very natural generalization (specific to
  so-called ``\css'' codes) is adequate for our purposes.
Along the way, we provide modified versions of the classical
  sharing protocols of \cite{CCD88}. The new property our protocols
  have is that dealers do not need to remember the randomness they use
  when constructing shares to distribute to other players. This allows
  them to replace a random choice of coins with the
  \emph{superposition} over all such choices.

\subsection{Previous Work}
\label{sec:prevwork}

\mypar{Classical {\sc mpc}}
Multi-party computing was introduced by Goldreich, Micali and
Wigderson \cite{GMW87}, who showed that \emph{under computational
  assumptions}, secure multi-party evaluation of any function was
possible tolerating any minority of cheating players, i.e. if and only
if $t <\frac n 2$.  If one assumes pairwise secure channels but no
computational assumptions, then one can compute any function securely
if and only if $t< n/3$ \cite{BGW88,CCD88}.  If one further assumes
the availability of a secure broadcast channel, then one can in fact
tolerate $t<n/2$, and no more (\cite{RB89,Beaver89,CDDHR99}).  All of
these protocols rely on verifiable secret sharing as a basic tool. Our
solution draws most heavily on the techniques of Chaum, Cr\'epeau and
Damg{\aa}rd \cite{CCD88}.

Beyond these basic protocols, much work has focused on finding proper
definitions of security, e.g. 
\cite{GL90,Beaver91,MR91,PW00,Can01}.
We adopt a simple definition based on the initial definitions of
Canetti.

\mypar{Quantum Secret Sharing}
Relatively little work exists on multi-party cryptographic protocols
with quantum data.  Secret sharing with a quantum secret was first
studied by Cleve et al.~\cite{CGL99}, who showed an equivalence with
quantum error-correcting codes (QECC). Their scheme is the basis of
our protocols.
%
%
%
Chau \cite{Chau00} deals with classical computations, but also
mentions the problem of verifiable quantum secret sharing as an open
question.

\mypar{Fault-tolerant Quantum Computing}
%
The goal of \ftqc\ is to tolerate
\emph{non-malicious} faults occurring within a single computer. One
assumes that at every stage in the computation, every qubit in the
circuit has some known probability $p$ of suffering a random error,
i.e. of becoming completely scrambled. Moreover, errors are assumed to
occur \emph{independently} of each other and of the data in the
computation.

One can view multi-party computation as fault-tol\-er\-ant computing
with a different error model, one that is suited to distributed
computing.  The \mpqc\ model is weaker in some respects since we
assume that errors will always occur in the same, limited number of
positions, i.e. errors will only occur in the systems of the $t$
corrupted players.
In other respects, the error model of \mpqc\ is stronger: in our
setting errors may be \emph{maliciously} coordinated.  In particular,
they will not be independently placed, and they may in fact depend on
the data of the computation---the adversaries will use any partial
information known about the other players' data, as well as
information about their own data, to attempt to corrupt the
computation. For example, several \ftqc\ algorithms rely on the fact
that at certain points in the computation, at most one error is likely
to occur. Such algorithms will fail when errors are placed
adversarially.
Techniques from \ftqc\ are nonetheless useful for multi-party
computing. We will draw most heavily on techniques due to Aharonov and
Ben-Or \cite{AB99}.

\subsection{Definitions and Model}
\label{sec:def}

In this paper, we use a simple simulation-based framework for proving
the security of quantum protocols, similar to early classical
definitions. We specify a task by giving a protocol for implementing
it in an ideal model where players have access to a trusted third
party $\ttp$. We prove a given protocol secure by showing a simulator
which translates any attack in the real-world protocol into an
(almost) equally successful attack in the ideal model.



We assume that every pair of participants is connected by perfect
(i.e.  authenticated, unjammable, secret) quantum and classical
channels, and that there is a classical authenticated broadcast
channel to which all players have access.  Because we will always
consider settings where $t < \frac n 2$, we can also assume that
players can perform
\emph{classical} multi-party computations securely \cite{CDDHR99}%
\footnote{In fact, even the assumption of a broadcast channel is not
  strictly necessary, since $t<\frac n 3$ in our setting.}.
The adversary is an arbitrary quantum algorithm (or family of
circuits) $\mathcal{A}$ (not necessarily polynomial time), and so the
security of our protocols does not rely on computational assumptions.

The real and ideal models, as well as the notion of security, are
specified more carefully in \cite{Sm01}.  In this paper, we use the
following informal specifications of the ideal protocols.  The real
protocols are secure if they succeed in simulating the ideal ones.

\mypar{Multi-party Quantum Computation}
\label{sec:mpc-def}
%
All players hand their
inputs to the $\ttp$, who runs the desired circuit and hands
back the outputs. Note that the only kind of cheating which is
possible is that cheaters may choose their own input. In particular,
cheaters cannot force the protocol to abort.


\mypar{Verifiable Quantum Secret Sharing}
\label{sec:vqss-def}
%
In the sharing phase, the dealer gives his secret system to the
trusted party. In the reconstruction phase, the $\ttp$ sends the
secret system to the reconstructor $R$. The only catch is that in the
ideal model, honest players should not learn the identity of $R$ until
after the first phase has finished (otherwise, $D$ could simply send
the secret state to $R$ in the first phase without violating the
definition). 

\subsection{Preliminaries}
\label{sec:prelim}

We present the notation necessary for reading the protocols and proofs
in this paper. For a more detailed explanation of the relevant
background, see \cite{Sm01} or a textbook such as \cite{NC00}.

We will work with $p$-dimensional quantum systems, for some prime
$p>n$. Such a system is called a qupit, and the ``computational''
basis states are labelled by elements in $F=\ze_p$. We will also be
working in the Fourier basis, which is given by the unitary
transformation $\F \ket{a} \mapsto \sum_{b} \omega^{ab}\ket{b}$.  A
basis for the operators on a qupit is given by the $p^2$ Pauli
operators $X^aZ^b$, where $X\ket a = \ket {a+1}$ , $Z\ket{a} =
\omega^a\ket a$, and $\omega = \exp(2\pi i/p)$. Tensor products of
these operators yield the Pauli basis for the set of operators on a
register of qupits. The weight of a tensor product operator is the
number of components in which it is not the identity $\mathbb{I}$.

\mypar{Quantum Codes} The error-correcting codes used in this paper
are quantum CSS codes.  These are defined via two classical linear
codes $V,W \subseteq \ze_p^n$ such that $\vp \subseteq W$. If we
denote $W^{(q)} = \mathrm{span}\{\ket{{\bf w}}:\ {\bf w} \in W\}$ for
a classical code $W$, then we can write the CSS code as $\C = \vq \cap
\F\wq$. Thus, $\C$ is the set of states of $n$ qubits which yield a
codeword of $V$ when measured in the computational basis and a
codeword of $W$ when measured in the Fourier basis.

Specifically, we will use quantum Reed-Solomon codes from \cite{AB99}.
We specify a quantum RS code by a single parameter $\delta< n/2$.
The classical Reed-Solomon code $V^{\delta}$ is the set of all vectors
$\mathbf{\hat{q}}=\left( q(1),q(2),\ldots,q(n)\right)$, where $q$ is
any univariate polynomial of degree at most $\delta$. The related
code $V^{\delta}_0$ is the subset of $V^\delta$ corresponding to
polynomials which interpolate to 0 at the point 0. That is: $ V^\delta
= \{ \mathbf{\hat q}:\ q \in F[x] :\ \deg(q) \leq \delta\} $ and
$V_0^\delta = \{ \mathbf{\hat q}:\ \deg(q) \leq \delta \mbox{ and
  }q(0)=0\} \subseteq V^\delta $.  The code $V^{\delta}$ has minimum
distance $d=n-\delta$, and an efficient error correction procedure.
Let $\delta'=n-\delta-1$. There are constants $d_1,...,d_n \in \ze_p$
such that the dual of the code $V^\delta$ is just the code
$V_0^{\delta'}$, rescaled by $d_i$ in the $\ith$ coordinate;
similarly, the dual of $V_0^{\delta}$ is a rescaled version of
$V^{\delta'}$. Denote these duals by $W_0^{\delta'}, W^{\delta'}$,
respectively.

The quantum code $\C^\delta$ for parameter $\delta$ is the \css\ code
obtained from codes $V=V^\delta$ and $W=W^{\delta'}$. It encodes a
single qupit, and has minimum distance $\delta+1$ (thus, it corrects
$t=\floor{\delta/2}$ errors). Moreover, errors can be corrected
efficiently, given the syndrome of a corrupted codeword, i.e. the
$V$ syndrome measured in the computational basis and the $W$ syndrome
measured in the Fourier basis.  

\an{Is the following too brief?}

\mypar{Transversal Operations}
A nice result from fault-tolerant computing \cite{AB99,Gpers} is that
one can in fact perform 
many operations on data encoded by a quantum RS code using only local
operations and classical information transmitted between the
components.
Consider the following gates:

\vspace{4pt}

\begin{narenum*}\setlength{\itemsep}{2pt}
\item \label{gate:x} Shift: 
  $X^c: \ket a \mapsto \ket{a+c}$,
\item SUM: $(c\mbox{-}X):
  \ket{a,b}\mapsto \ket{a,a+b}$,
\item Scalar multiplication: $0\neq c\in F$,
  $S_c: \ket{a}\mapsto \ket{ac}$,
\item \label{gate:z} Phase Shift: 
  $Z^c: \ket{a}\mapsto w^{ca}\ket{a}$,
\item Fourier Transform: $\F_r: \ket{a} \mapsto
  \frac{1}{\sqrt{p}}\sum_{b\in F} w^{rab}\ket{b}$,
\item Toffoli (Multiplication):
$\ket{a}\ket{b}\ket{c}\mapsto \ket{a}\ket{b}\ket{c+ab}$.
\end{narenum*}

\vspace{4pt}

These gates are universal~\cite{AB99}, in the sense that a sequence of
these gates can approximate any unitary operation with arbitrary
accuracy.  Beyond these, in order to simulate arbitrary quantum
circuits one should also be able to introduce qupits in some known
state (say $\ket{0}$), as well as to discard qupits.  For any \css\
code, the gates \ref{gate:x} through \ref{gate:z} from the set above
can be implemented \emph{transversally}, that is using only local
operations which affect the same component of two codewords.
Measurement and the remaining two operations can be performed almost
transversally.

\mypar{Measurement} 
For a quantum RS code, measuring each component of
the encoding of $\ket{s}$ yields a vector ${\bf \hat
  q}=(q(1),...,q(n))$ where $q(0)=s$.  This operation is not quite
transversal since after the qupit-wise measurement, the classical
information must be gathered together in order to extract the
measurement result. Nonetheless, it can tolerate arbitrary corrupton
of $\delta/2$ of the positions in the codeword if classical
error correction is first applied to the vector of measurement
results.

\mypar{Fourier and Toffoli gates}
For \css\ codes, applying the Fourier transform transversally maps
data encoded with the codes $V,W$ to the Fourier transform of that
data, encoded with the dual code $\TC$ defined via the codes $W,V$.
For quantum RS codes, rescaling each component of the dual code of
$\C^\delta$ produces the code $\C^{\delta'}$. This allows one to
perform the map $\mathcal{E}_{\C^\delta} \ket\psi \mapsto
\mathcal{E}_{\C^{\delta'}} \paren{ \F \ket\psi}$, where $\E_{\C}$ is
the encoding map for a code $\C$.

When $n=2\delta+1$, we have $\delta'=\delta$, so the Fourier transform
is in fact transversal, but the Toffoli gate is difficult to perform.

When $n=3\delta+1$, neither the Fourier transform nor the Toffoli gate
is transversal, but they can both be reduced to {\em degree reduction}
via transversal operations~\cite{AB99}. Degree reduction maps an
arbitrary state $\ket \psi$ encoded using $C^{\delta'}$ to $\ket \psi$
encoded with $C^{\delta}$.

The circuit we use for degree reduction is due to Gottesman and
Bennett \cite{Gpers}.  We start with one block encoded using
$\C^{\delta'}$ (system $\mathcal{H}_1$), and an ancilla block in the
state $\mathcal{E}_{\C^{\delta}} \left(\sum \ket{a}\right)$ (system
$\mathcal{H}_2$).  Perform a SUM gate from $\mathcal{H}_2$ to
$\mathcal{H}_1$ (this can be done transversally by a property of the
codes $\C^{\delta}$).  Measure $\mathcal{H}_1$ in the computational
basis, obtaining $b$, and apply $X^{b} S_{-1}$ to $\mathcal{H}_2$.
The system $\mathcal{H}_2$ now contains the data, encoded using
$\C^{\delta}$.  This entire procedure can be performed transversally
except for the measurement step.  However, as noted above, measurement
requires only classical communication between the components.

\section{Neighborhoods of Quantum\\ Codes}
\label{sec:nbhds}

One of the ideas behind classical multi-party computing protocols is
to ensure that data is encoded in a state that remains ``close'' to a
codeword, differing only on those positions held by cheaters (call
that set $B$).
%
For classical codes, ``close'' means that the real word $\mathbf v$
should differ from a codeword only on $B$, so that any errors
introduced by cheaters are correctable.
%
%
For a code $W$, let the $B$-neighborhood $W_B$ be the set of vectors
differing from a codeword of $W$ by positions in $B$, i.e., $$W_B =
\set{\mathbf{v} \ :\ \exists \mathbf{w} \in W\ \mbox{s.t.}\ 
  \supp(\mathbf{v-w})\in B}.$$
%
%
Equivalently, one can define $W_B$ as the set of words obtained by
distributing a (correct) codeword to all players, and then having all
players send their shares to some (honest) reconstructor $R$.


For quantum codes, there is more than one natural definition of the
neighborhood corresponding to a set $B$ of positions. Let
$\set{1,...,n}$ be partitioned according to two sets $A,B$.
We say a mixed state $\rho'$ is ``in'' $\C$ if all states in the
mixture lie in $\C$, i.e.  $\tr(P_\C \rho')=1$ where $P_\C$ is the
projector onto $\C$. We consider three definitions of a
``$B$-neighborhood'' of a \css\ code $\C$. Let $\rho$ be an arbitrary
state of the coding space.

\vspace{4pt}
\begin{narenum*}\setlength{\itemsep}{2pt}
\item\label{nbhd1} $\rho$ differs from a state in $\C$ only by some
  quantum superoperator $\mathcal{O}$ acting only on  $B$:\\
  $ N_B(\C) = \set{\rho : \exists \rho'\ \mbox{in}\ \C, \exists
    \mathcal{O}  \mbox{s.t.}\ \rho = \mathcal{O}(\rho') }$.
  
\item \label{nbhd2} $\rho$ cannot be distinguished from a state in
  $\C$ by looking only at positions in $A$.\\
  $ST_B(\C) = \set{\rho : \exists \rho'\ \mbox{in}\ \C\ \mbox{s.t.}\ 
    \tr_B(\rho) = \tr_B(\rho')}$.   
\item \label{nbhd3} Specifically for \css\ codes, one can require that
  the state $\rho$ pass checks on $A$ in both bases, i.e.  that
  measuring either the $V_B$ syndrome in the computational basis, or
  the $W_B$ syndrome in the Fourier basis, yields 0.  The set of
  states which pass this test is: $\C_B = \vbq \cap \fn \wbq.$
\end{narenum*}
\vspace{4pt}

In general, these notions form a strict hierarchy: \\
$N_B(\C) \subsetneq ST_B(\C)\subsetneq\C_B$. Only notion (\ref{nbhd3})
is always a subspace (see \cite{Sm01} for details).

%

In the analysis of quantum error correction and fault-tolerance
schemes, it is sufficient to consider notion (\ref{nbhd1}), for two
reasons. On one hand, one starts from a correctly encoded state. On
the other hand, the errors introduced by the environment will be
independent of the encoded data (and in fact they must be for
error correction to be possible at all in that context).

In our setting, however, we cannot make such assumptions. The
cheaters might possess states which are entangled with the data in the
computation, and so the errors they introduce will not be independent
of that data. Instead, we show that our verifiable sharing protocol
guarantees a condition similar to notion (\ref{nbhd3}) (see
\lemref{l2-sound}).  In order to provide some intuition for the proofs
of \secref{2ls}, we characterize notion (3) below.

\mypar{Well-Definedness of Decoding for $\C_B$} 
The set $\C_B$ is a subspace, since it is defined in terms of
measurement outcomes. More particularly, it is 
spanned by the states of $N_B(\C)$:

\begin{lemma}\label{lem:charact}
  If $\rho$ is in $\C_B=\vbq \cap \fn \wbq$, then we can write
  $\rho=\sum_i p_i\ket{\psi_i}\bra{\psi_i}$, where $\ket{\psi_i} =
  \sum_j c_{ij}E_j\ket{\phi_{ij}}$, the $E_j$ are Pauli operators on
  $B$ and $\ket{\phi_{ij}}\in {\cal C}$.
\end{lemma}

\begin{proof}
  To check if $\rho$ is in $\C_B$, we measure the $V_B$ syndrome in
  the computational basis and the $W_B$ syndrome in the Fourier basis.
  However, the distribution on this outcome measurement will not
  change if we first measure the $V$ and $W$ syndromes, i.e.  if we
  first make a measurement which projects $\rho$ into one of the
  subspaces $E_j\C$ (i.e. $\rho$ maps to $\rho'=P_j \rho P_j$ with
  probability $\tr\paren{P_j \rho}$, where $P_j$ is the projector for
  the space $E_j\mathcal{C}$).
  
  The new state $\rho'$ lies completely in one of the spaces $E_j\C$.
  However, $E_j\C$ is either contained in $\C_B$ (if there is an
  operator equivalent to $E_j$ which acts only on $B$) or
  \emph{orthogonal} to $\C_B$ (if no such operator exists).

  Thus
  $\tr\paren{P_j\rho}=0$ for all $E_j$ which act on more than
  $B$. Hence $\rho$ is a mixture of states $\ket{\psi_i}$ each of
  which is a linear combination of elements of the spaces
  $\set{E_j\C}$, where $E_j$ acts only on $B$.
\end{proof}

This has a useful corollary, namely that decoding is well-defined for
states in $\C_B$. Formally, there are two natural ``reconstruction
operators'' for extracting the secret out of a state which has been
shared among several players. Suppose that $\C$ has distance $d>2t+1$
and $|B|\leq t$.
%
First, ${\cal D}$ is the decoding operator for the error-correcting
code $\mathcal{C}$, which would be applied by an honest player holding
all of the shares. For any operator $E_j$ of weight less than $t$ and
for any state $\E \ket{\phi}$ in $\mathcal{C}$, we have
$\mathcal{D}E_j\E\ket{\phi} = \ket{\phi} \otimes \ket{j}$ (i.e.  the
error is not only corrected but also identified). It will then discard
the system containing the syndrome information $\ket{j}$.
Second, $\ri$ is the ``ideal recovery operator'', defined by
identifying the set $B$ of cheaters and applying the simple
interpolation circuit to any set of $n-2t$ good players' positions
(this corresponds to erasure recovery).

\begin{prop}\label{prop:l1-well-defined}
  For any state $\rho$ in $\mathcal{C}_B$, the state $\ri (\rho)$ is
  well-defined and is equal to $\mathcal{D}(\rho)$.
\end{prop}

Our protocols guarantee conditions similar to $\mathcal{C}_B$, and 
well-definedness is essential for proving simulatability.

\begin{proof}
  Consider a particular basis state $E_j\E \ket a$. The decoding
  operator $\mathcal{D}$ will produce the state $\ket a\ket j$, since
  errors of weight at most $t$ can be identified uniquely.  The ideal
  operator $\ri$ will extract the encoded state $\ket{a}$. Without
  loss of generality, the ideal recovery operator will replace $\ket
  a$ with $\ket 0$, the final output $\ket a \otimes E_j\E\ket{0}$.
  
  In both cases, the output can be written as $\ket{a}$ tensored with
  some ancilla whose state depends only on the syndrome $j$ (and which
  identifies $j$ uniquely).  Once that state is traced out, the
  outputs of both operators will be identical. Another way to see this
  is that the ideal operator can simulate the real operator: one can
  go from the output of the ideal operator to that of the real
  operator by applying a transformation which only affects the
  ancilla. For a state $\rho$ expressed as in \lemref{charact}, the
  final outcome will be $\rho' = \sum_{ij} p_i |c_{ij}|^2
  \proj{\phi_{ij}}$.
\end{proof}


\section{A Two Level VQSS Protocol}
\label{sec:2ls}

In this section we define a two-tiered protocol for \vqss. It is based
on the \vqss\ protocols of \cite{CCD88} as well as on the literature
on quantum fault-tolerance and error correction, most notably on
\cite{AB99}.  Detailed proofs for the claims of this section are in
\cite{Sm01}. However, some intuition is given by the proofs of 
\secref{nbhds}.


\subsection{Sharing Shares: 2-GOOD Trees}
\label{sec:hier}

In the \cvss\ protocol of \cite{CCD88}, the dealer $D$ takes his
secret, splits it into $n$ shares and gives the $\ith$ component to
player $i$. Player $i$ then shares this secret by splitting it into
$n$ shares and giving the $\jth$ share to player $j$. Thus, there are
$n^2$ total shares, which can be thought of as the leaves of a tree
with depth 2 and fan-out $n$: each leaf is a share; the $\ith$ branch
corresponds to the shares created by player $i$, and the root
corresponds to the initial shares created by the dealer.  Player $j$
holds the $\jth$ leaf in each branch of this tree.
We will run a cut-and-choose protocol 
in order to guarantee some kind of
consistency of the distributed shares.

During the protocol we accumulate $n+1$ sets of apparent cheaters:
one set $B$ for the dealer (this corresponds to a set of branches
emanating from the root), and one set $B_i$ for each player $i$ (this
corresponds to a subset of the leaves in branch $i$).  These sets all
have size at most $t$.
%
%
At the end of the protocol, we want to guarantee certain invariants.
Say $V$ has minimum distance $>2t$, and each codeword corresponds to a
single value $a \in \ze_p$.
\begin{definition}[$\tgood$ trees]\label{def:tgood} 
  We say a tree of $n^2$ field elements is $\tgood$ with respect to
  the code $V$ and the sets $B,B_1,...,B_n$ if:
  \begin{narenum*}
  \item For each $i \not \in C$ (i.e., corresponding to an honest 
    player), we have $B_i\subseteq C$, i.e. all apparent cheaters are 
    real cheaters.

  \item For each branch $i \not \in B$, the shares held by the honest
    players \emph{not in} $B_i$ should all be consistent 
    with some codeword in $V$, 
    i.e. the vector of all shares
    should be in $V_{B_i \cup C}$, where $C$ is the set of cheating
    players.

    N.B.: Because there are at most $t$ players in $B_i$ and at most
    $t$ cheaters, there are at least $d+1\leq n-2t$ honest players
    remaining, and so the polynomial above is uniquely defined. This
    guarantees that for each branch $i \not \in B$, there is a unique
    value $a_i\in F$ which is obtained by interpolating the shares of
    the honest players not in $B_i$.

  \item For $i\not \in B$, the values $a_i$ defined by the previous
    property are all consistent with a codeword of $V$ (i.e. the
    vector $(a_1,...,a_n)$ is in $V_B$).
  \end{narenum*}
  We will abbreviate this as $\tgood_V$, when the sets
  $B$, $B_1$,$...$,$B_n$ are clear from the context.
\end{definition}

\subsection{VQSS Protocol}
\label{sec:vqss}

The 
{\sc Vqss} protocol is described in Protocols \ref{prot:vqss} and
\ref{prot:qr}.  Intuitively, it guarantees that a tree of quantum
shares would yield a $\tgood$ tree of classical values if measured in
either the computational basis or the Fourier basis.  We use the codes
$V=V^{\delta}=V^{\delta'}$ and $W=W^{\delta}=W^{\delta'}$, with
$n=4t+1,\delta=\delta'=2t$, although there is in fact no need to do
this: the protocol will work for any \css\ code with distance at least
$2t+1$, so long as the codes $V,W$ are efficiently decodable.

\newlength{\adamwidth}
\setlength{\adamwidth}{460pt}

 \begin{figure*}[t]
     \fbox{\parbox[t]{\adamwidth}{
{
\begin{protocol}[{\vqss---Sharing Phase}] \label{prot:vqss}
  Dealer $D$ gets as input a quantum system $S$ to share. 
  \begin{naritem}
  \item \textbf{Sharing:}
    \begin{narenum}
    \item The dealer $D$ prepares $(k+1)^2$ systems of $n$ qupits
      each, called $S_{\ell,m}$ (for $\ell=0,...,k$ and $m=0,...,k$):
      \begin{narenum}
      \item Encodes $S$ using ${\cal C}$ in $S_{0,0}$.
      \item Prepares $k$ systems $S_{0,1},...,S_{0,k}$ in the state
        $\sum_{a \in F} \E_\C \ket{a} = \sum_{v \in V} \ket{v}$.
      \item Prepares $k(k+1)$ systems $S_{\ell,m}$, for $\ell=1,...,k$
        and $m=0,...,k$, each in the state $\ket{\bar 0}= \sum_{v \in
          V_0} \ket{v}$.
      \item For each of the $(k+1)^2$ systems $S_{\ell,m}$, $D$ sends
        the $\ith$ component (denoted $S_{\ell,m}^{(i)}$) to player $i$.
      \end{narenum}
    \item Each player $i$, for each $\ell,m=0,...k$:
      \begin{narenum}
      \item Encodes the received system $S_{\ell,m}^{(i)}$ using
        ${\cal C}$ into an $n$ qupit system $S_{\ell,m,i}$.
      \item Sends the $\jth$ component $S_{\ell,m,i}^{(j)}$ to player
        $j$.
      \end{narenum}
    \end{narenum}
  \item \textbf{Verification:}
    \begin{narenum}
    \item \label{step:l2-vcomp} Get public random values $b_1,...,b_k
      \in_R F$. For each $\ell=0,...,k$, $m=1,...,k$, each player $j$:
      \begin{narenum}
      \item Applies the SUM gate $(c\mbox{-}X^{b_m})$
        to his shares of the systems $S_{\ell,0,i}$ and
        $S_{\ell,m,i}$.
      \item Measures his share of $S_{\ell,m,i}$ and broadcasts the
        result (i.e. each player broadcasts $k(k+1)n$ values).
      \item \label{step:l2-update}
        For each $i\in \set{1,...,n}$, players update the set $B_i$
        based on the broadcast values: there are $(k+1)kn$ broadcast
        words ${\bf w}_{\ell,m,i}$. Applying classical decoding to
        each of these yields min-weight error vectors ${\bf
          e}_{\ell,m,i}$ with supports $B_{\ell,m,i}$. Set $B_i =
        \cup_{\ell,m} B_{\ell,m,i}$.  If there are too many errors,
         add $i$ to the global set $B$.

      \item Furthermore, players do the same at the root level: for
        all $i \not \in B$, there is an interpolated value $a_i$ which
        corresponds to the decoded codeword from the previous step.
        Players also decode the codeword $(a_1,...,a_n)$ and update
        $B$ accordingly (i.e. by adding any positions where errors
        occur to $B$).
    \end{narenum}
  \item \label{step:l2-rotate} All players apply the Fourier
    transform $\F$ to their shares.
  \item \label{step:l2-vfour} Get public random values
    $b_1',...,b_k' \in_R F$. For $\ell=1,...,k$, each player $j$:
    \begin{narenum}
    \item Applies the SUM gate $(c\mbox{-}X^{b_\ell'})$
      to his shares of the systems $S_{0,0,i}$ and $S_{\ell,0,i}$.
    \item Measures his share of $S_{\ell,0,i}$ and broadcasts the
      result (i.e. each player broadcasts $kn$ values).
    \item For each $i\in \set{1,...,n}$, players update $B_i$ and $B$
      based on the broadcast values (as in \stepref{l2-update}).
      
      [Note: the sets $B$ and $B_1,...,B_n$ are cumulative throughout
      the protocol.]
    \end{narenum}
  \item \label{step:l2-frotate} All players apply the inverse
    transform $\F^{-1}$ to their shares of $S_{0,0}$.
  \end{narenum}
  
\item The remaining shares (i.e. the components of the $n$ systems
  $S_{0,0,i}$) form the sharing of the state $\rho$.
\end{naritem}
\end{protocol}}}}


\vspace{8pt}

     \fbox{\parbox[t]{\adamwidth}
{{
  \begin{protocol}[{\vqss---Reconstruction Phase}]\label{prot:qr} 
    Player $j$ sends his share of each of the systems $S_{0,0,i}$ to
    the receiver $R$, who runs the following decoding algorithm:
    \begin{narenum*}
    \item For each branch $i$: Determine if there is a set $\tilde
      B_i$ such that $B_i \subseteq \tilde B_i$, $|\tilde B_i|
      \leq t$ and the shares of $S_{0,0,i}$ lie in ${\cal C}_{\tilde
        B_i}$.  \\
      If \emph{not}, add $i$ to $B$.
      Otherwise, correct errors on $\tilde B_i$ and decode to obtain a
      system $S_i'$.
    \item Apply interpolation to any set of $n-2t$ points not in $B$.
      Output the result $S'$.
    \end{narenum*}
  \end{protocol}}}}
 
 \end{figure*}

The protocol can be tweaked for efficiency. The final protocol takes
three rounds. 
Each player sends and receives $O(n + \log \frac 1 \epsilon)$ qubits,
and the broadcast channel is used $O\paren{n(n + \log \frac 1
  \epsilon)}$ times overall, where $\epsilon$ is the soundness error
of the protocol (this requires setting $k= n + \log (\frac 1
\epsilon)$).

Why is this a secure \vqss\ protocol? We want to show that the protocol
is equivalent to the ``ideal model'', where at sharing time the dealer
sends his secret system $S$ to a trusted outside party, and at reveal
time the trusted party sends $S$ to the designated receiver. To do
that, we will use two main technical claims.


\mypar{Soundness}
\label{sec:l2-sound}
We must show that at the end of the protocol, if the dealer passes all
tests then there is an well-defined ``shared state'' which will be
recovered by the dealer. To do so, we guarantee a property similar to
$\C_C$ (notion (\ref{nbhd3}) of \secref{nbhds}).

\begin{lemma}\label{lem:l2-sound}
  The system has high fidelity to the following statement: ``Either
  the dealer is caught (i.e. $|B| > t$) or measuring all shares in the
  computational (resp.  Fourier) basis would yield a $\tgood$ tree
  with respect to the code $V$ (resp. $W$).''
\end{lemma}

Proof of this is via a ``quantum-to-classical'' reduction, similar to
that of \cite{LC99kd}. First, checks in the computational and Fourier
bases don't interfere with each other, since they commute for CSS
codes.  Second, in a given basis, we can assume w.l.o.g. that all
ancillae are first measured in that basis, reducing to a classical
analysis similar to \cite{CCD88}.

\mypar{Ideal Reconstruction}
In order to prove soundness carefully, we define an \emph{ideal
  interpolation} circuit $\ri$ for $\tgood$ trees: pick the
first $n-2t$ honest players not in $B$, say $i_1,...,i_{n-2t}$. For
each $i_j$, pick $n-2t$ honest players not in $B_{i_j}$ and apply the
normal interpolation circuit (i.e.  erasure-recovery circuit) for the
code to their shares to get some qupit $R_{i_j}$. 
Applying the interpolation circuit again, we
extract some system $S$ which we take to be the output of the ideal
interpolation. 

The \emph{real} recovery operator $\mathcal{D}$ is given by
\protref{qr}.
The following lemma then applies, following essentially from
\propref{l1-well-defined}.

\begin{lemma}\label{lem:l2-well-defined}
  Given a tree of qupits which is $\tgood$ in both bases, the outputs
  of $\ri$ and $\D$ are the same. In particular, this means that no
  changes made by cheaters to their shares 
  can affect the outcome of $\D$.
\end{lemma}


Lemmas \ref{lem:l2-sound} and \ref{lem:l2-well-defined} show that
there is essentially a unique state which will be recovered in the
reconstruction phase when the receiver $R$ is honest.



\mypar{Completeness} 
\label{sec:l2-complete}
As discussed earlier, the protocol is considered complete if when the
dealer is honest, the state that is recovered by an honest
reconstructor is exactly the dealer's input state. The key property is
that  \emph{
  when the dealer $D$ is honest, the effect of the verification phase
  on the shares which never pass through cheaters' hands is the
  identity.}

Consider the case where the dealer's input is a pure state
$\ket{\psi}$. On one hand, we can see by inspection that an honest
dealer will always pass the protocol. Moreover, since the shares that
only go through honest players' hands remain unchanged, it must be
that if some state is reconstructed, then that state is 
$\ket{\psi}$, since the ideal reconstruction operator uses only those
shares. Finally, we know that since the dealer passed the protocol the
overall tree must be $\tgood$ in both bases, and so some value will be
reconstructed. Thus, on input 
$\ket\psi$, an honest reconstructor will reconstruct $\ket\psi$. We have proved:

\begin{lemma}\label{purestate}
  If $D$ and $R$ are honest, and the dealer's input is a pure state
  $\ket\psi$, then $R$ will reconstruct a state $\rho$ with fidelity
  $1-2^{-\Omega(k)}$ to the state $\ket\psi$.
\end{lemma}

Not surprisingly, this lemma also guarantees the privacy of the
dealer's input. By a strong form of the no cloning theorem,
any information the cheaters could obtain would cause some
disturbance, at least for a subset of the inputs. Thus, the protocol
is in fact also private.


\mypar{Simulatability}
The claims above show that the protocol satisfies an intuitive notion
of security. In this section we sketch a proof that the protocol
satisfies a more formal notion of security: it is equivalent to a
simple ideal model protocol. The equivalence is \emph{statistical},
that is the outputs of the real and ideal protocols may not be
identical but will have very high fidelity to one another.



\begin{theorem}\label{thm:vqss}
  Protocols \ref{prot:vqss} and \ref{prot:qr} are a statistically secure 
  \vqss\ scheme.
\end{theorem}



The ideal protocol is sketched in \secref{def}. 
To show equivalence, 
we will give a transformation that takes an adversary $\A_1$ for our
protocol and turns it into an adversary $\A_2$ for the ideal protocol.
To give the transformation, we exhibit a simulator $\Sim$ which acts
as an intermediary between $\A_1$ and the ideal protocol, making
$\A_1$ believe that it is experiencing the real protocol.

The idea is that the simulator will simulate the regular \vqss\ 
protocol either on input provided by a cheating dealer or on bogus
data $\ket{0}$, and then extract and/or change the shared state as
needed.


\begin{figure*}[bt]
    \fbox{ \parbox[t] { \adamwidth }{ 
      \begin{algm} \label{alg:vqss-sim} Simulation for \vqss\ (\protref{vqss})
        \begin{naritem}
        \item {\bf Sharing/Verification phase} {~}
              \begin{naritem}
              \item If $D$ is a cheater, $\Sim$ must extract some
                system to send to $\ttp$:
                \begin{narenum}
                \item Run Sharing and Verification phases of
                  \protref{vqss}, simulating honest players. If $D$ is
                  caught cheating, send ``I am cheating'' from $D$ to
                  $\ttp$.
                \item Choose $n-2t$ honest players not in $B$ and
                  apply ideal interpolation circuit to extract a
                  system $S$.
                \item Send $S$ to $\ttp$.
                \end{narenum}
              \item If $D$ is honest, $\Sim$ does not need to send
                anything to $\ttp$, but must still simulate the sharing
                protocol.
                \begin{narenum}
                \item Simulate an execution of the Sharing and
                  Verification phases of \protref{vqss}, using $\ket
                  0$ as the input for the simulated dealer $D'$.
                \item Choose $n-2t$ honest players (they will
                  automatically not be in $B$ since they are honest)
                  and apply the ideal interpolation circuit to extract
                  the state $\ket 0$.
                \item The honest $D$ will send a system $S$ to $\ttp$.
                \end{narenum}
              \end{naritem}
              \textbf{Note:} Regardless of whether $D$ is honest or
              not, at the end of the sharing phase of the simulation,
              the joint state of the players' shares is a tree that is
              (essentially) $\tgood$ in both bases, and to which the
              ideal interpolation operator has been applied.  Let $I$
              be the set of $n-2t$ honest players (not in $B$ or $C$)
              who were used for interpolation.
            \item {\bf Reconstruction phase} ~
              \begin{naritem}
              \item If $R$ is a cheater, $\Sim$ receives the system
                $S$ from $\ttp$. $\Sim$ runs the interpolation circuit
                backwards on the positions in $I$, with $S$ in the
                position of the secret. $\Sim$ sends the resulting
                shares to $R$.
              \item If $R$ is honest, the cheaters send their
                corrupted shares to $\Sim$. These are discarded by
                $\Sim$.
              \end{naritem}
              In both cases, $\Sim$ outputs the final state of $\A_1$
              as the adversary's final state.
            \end{naritem}
          \end{algm}
        }}
\end{figure*}


We give a sketch of the simulation procedure in \algref{vqss-sim}. Why
does this simulation work?
\vspace{2pt}
\begin{narenum*}
\item When $D$ is cheating:
  \begin{narenum}
  \item When $R$ is cheating, the simulation is trivially faithful,
    since there is \emph{no difference} between the simulation and the
    real protocol: $\Sim$ runs the normal sharing protocol, then runs
    the interpolation circuit, sending the result to $\ttp$. In the
    reconstruction phase, $\Sim$ gets the same state back from $\ttp$,
    and runs the interpolation circuit in reverse. Thus, the two
    executions of the interpolation circuit cancel out.
  \item When $R$ is honest, the faithfulness of the simulation comes
    from \lemref{l2-well-defined}: in the real protocol, $R$ outputs
    the result of the regular decoding operator. In the simulation,
    $R$ gets the output of the ideal interpolation. Since the shared
    state has high fidelity to a $\tgood$ tree (by \lemref{l2-sound}),
    the outputs will be essentially identical in both settings
    (i.e. they will have high fidelity).
  \end{narenum}

\item When $D$ is honest:
  \begin{narenum}
  \item When $R$ is also honest, the faithfulness of the simulation
    follows from the completeness and privacy properties of the real
    protocol.  Privacy implies that the adversary $\A_1$ cannot tell
    that it is actually participating in a sharing of $\ket 0$ rather
    than the dealer's state, and completeness means that $R$ in the
    real protocol gets a state with high fidelity to that received by
    $R$ in the ideal protocol.
  \item When $R$ is a cheater, $\Sim$ does not get $S$ from $\ttp$
    until the reconstruction phase.  Then he applies the ideal
    interpolation circuit to extract the $\ket 0$ state used during
    the verification phase, swaps $S$ with $\ket 0$, then runs the
    ideal interpolation circuit backwards.  Since the ideal
    interpolation circuit only acts on shares of the honest players,
    $\Sim$ is capable of performing these operations without tipping
    off $\A_1$ to the fact that it is in a simulation.  By the
    completeness property of the real protocol and the no-cloning
    theorem, the residual state left over after the ideal
    interpolation circuit (i.e., the state of the cheaters) has almost
    no correlation to the data shared using the circuit, so swapping
    in $S$ and running the circuit backwards gives us a state with
    high fidelity to the state that would have resulted from sharing
    $S$ directly with the same $\A_1$.  Thus, the simulation is
    faithful in this case as well.
 \end{narenum}
\end{narenum*}

We have essentially proved \thmref{vqss}.

\subsection{Additional Properties}
\label{sec:l2p}

Two-level sharings produced by the same dealer (using the protocol
above) have some additional properties, which will be useful for
multi-party computation. First of all, notice that there is no problem
in tracking the sets $B,B_1,...,B_n$ incrementally across various
invocations of the protocol for the same dealer, and so we assume
below that these sets are the same for different sharings from the
same dealer.
\vspace{2pt}
\begin{narenum*}
\item Some operations can be applied transversally to valid
  sharings. Applying the linear operation $(x,y)\mapsto(x,y+bx)$
  (denoted $c\mbox{-}X^b$) to all shares of two sharings effectively
  applies $c\mbox{-}X^b$ to the shared states. Similarly, applying the
  Fourier rotation transversally changes the sharing to the dual code
  and applies a logical Fourier rotation. 
  Finally, measuring all shares of a valid sharing in the
  computational basis and applying classical decoding yields the same
  result as measuring the shared state. Thus, players can measure
  without exchanging quantum information.
\item \label{rem:share0} The dealer can additionally use the protocol to
  prove to all players that the system he is sharing is exactly the
  state $\ket 0$: the ancillas he uses in this case will all be
  sharings of $\ket 0$ (instead of $\sum \ket a$). The verification
  step is the same as before, except now players verify that the
  reconstructed codeword at the top level interpolates to 0.
  Similarly, the dealer can prove that he is sharing a state $\sum_a 
  \ket a$. 
  This will be useful for sharing ancillas in the \mpqc\ protocol.
\end{narenum*}


\section{Lower Bound for VQSS}
\label{sec:imposs}

\begin{lemma}\label{lem:no4}
  No 4-player \vqss\ scheme tolerates one cheater.
\end{lemma}

\begin{proof} 
  Suppose such a scheme exists. Consider a run of the protocol in
  which all players behave perfectly honestly until the end of the
  sharing phase, at which point one (unknown) player introduces an
  arbitrary error.  However, an honest ``receiver'' Ruth, given access
  to the state of all players, must still be able to recover the
  shared state.  Thus, the joint state of all players constitutes a
  four-component {\sc qecc} correcting one error.  However, no such
  code exists, not even a mixed-state one, by the quantum Singleton
  bound~\cite{KL97}.
\end{proof}


The optimality of our \vqss\ scheme is an immediate corollary, since
any protocol tolerating $n/4$ cheaters could be used to construct a
four-person protocol tolerating one cheater by having each participant
simulate $n/4$ players in the original protocol:

\begin{theorem}
  No \vqss\ scheme for $n$ players exists which tolerates
  all coalitions of $\lceil n/4\rceil$ cheaters.
\end{theorem}

Note that we have only proved the impossibility of \emph{perfect}
\vqss\ protocols. However, the quantum Singleton bound still holds
when exact equality is replaced by approximate correctness, and so in
fact even statistical \vqss\ schemes are impossible when $t\geq n/4$.


\section{Multi-party Computation}
\label{sec:mpc}

In this section we show how to use the \vqss\ protocol of the previous
section to construct a multi-party quantum computing scheme.
First, we give a modified \vqss\ protocol. At the end of the protocol,
all players hold a single qupit. With high fidelity, either the dealer
will be caught cheating or the shares of all honest players will be
consistent in both the computational and Fourier bases, i.e. there is
no set $B$ of ``apparent cheaters''.  We then apply fault-tolerant
techniques to achieve secure distributed computation.

\subsection{Top-Level Sharing Protocol}
\label{sec:l3p}

We will now restrict attention to protocols tolerating $t<n/6$
cheaters, instead of $t<n/4$ cheaters as before.
Thus,
we take $n=6t+1$ for simplicity, and as before we set $\delta=2t$
(thus $\delta'=4t$). We will work with the \css\ code $\C$ given by
$V=V^\delta$ and $W=W^{\delta'}$. Recall that this is the \css\ code
for which there exist nearly-transversal fault-tolerant procedures
(\secref{prelim}). Our goal is to share a state so that at the end all
shares of honest players lie in ${\cal C}_C = V_C^{(q)} \cap \fn
W_C^{(q)}$.

The new scheme is given in \protref{l3-vqss}. The idea is that the
previous \vqss\ scheme allows distributed computation of linear gates
and Fourier transforms on states shared by the same dealer. It also
allows verifying that a given shared state is either $\ket 0$ or $\sum
\ket a$. The players will use this to perform a distributed
computation of the encoding gate for the code ${\cal C}$.  Thus, the
dealer will share the secret system $S$, as well as $\delta$ states
$\sum \ket a$ and $n-\delta-1$ states $\ket 0$. Players then apply the
(linear) encoding gate, and each player gets sent all shares of his
component of the output. As before, the main lemmas are soundness and
completeness of the protocol:


\begin{lemma}[Soundness]
  At the end of the sharing phase, the system has high fidelity to
  ``either the dealer is caught or the players' shares $S_1...S_n$ 
  lie in ${\cal C}_C$''.
\end{lemma}
\begin{lemma}[Completeness]
  When $D$ is honest, on pure state input $\ket{\psi}$, the shared
  state will have high fidelity to $\myspan\set{\E\ket{\psi}}_C$ (i.e.
  will differ from $\E\ket\psi$ only on the cheaters' shares).
\end{lemma}

Note the dealer can also prove that he has shared a
$\ket 0$ state (by showing that his input is $\ket 0$).

\begin{figure*}[tb] 
    \fbox{\parbox[t]{\adamwidth}{
\begin{protocol}[{Top-Level Sharing}] \label{prot:l3-vqss}
  Dealer $D$ takes as input a qupit $S$ to share.
\begin{itemize}
\item
  \begin{narenum}
  \item \textbf{(Distribution)} The dealer $D$:
    \begin{narenum}
    \item Runs the level 2 \vqss\ protocol on input $S$.
    \item For $i= 1,...,\delta$: 
      Runs level 2 sharing protocol to share state $\sum_a\ket a$
      (see \remref{share0} in \secref{l2p})
    \item For $i=1,...,n-\delta-1$: 
      Runs level 2 sharing protocol to share state $\ket 0$
      (see \remref{share0} in \secref{l2p})
    \end{narenum}
    Denote the $n$ shared systems by $S_1,...,S_n$ (i.e. $S_1$
    corresponds to $S$, $S_2,...,S_{\delta+1}$ correspond to
    $\sum_a\ket{a}$ and $S_{\delta+2},...,S_n$ correspond to $\ket{0}$).
    Note that each $S_i$ is a two-level tree, and thus corresponds to
    $n$ components in the hands of each player.
    
  \item \label{step:l3-comput} \textbf{(Computation)} Collectively,
    the players apply the Vandermonde matrix to their shares of 
    $S_1,...,S_n$. 
    (If $D$ is honest then system $S_i$  encodes the $i$-th
    component of an encoding of the input $S$).

  \item \label{step:l3roll-back} For each $i$, all players send their
    shares of $S_i$ to player $i$, who decodes them (as per
    \protref{qr}).
  \end{narenum}
\item \textbf{Quantum Reconstruction} Input to each player $i$ is the
  share $S_i$ and the identity of the receiver $R$.
  \begin{narenum}
  \item Each player $i$ sends his share $S_i$ to $R$.
  \item $R$ outputs ${\cal D}(S_1,...,S_n)$ and discards any ancillas
    ($\D$ is the decoding algorithm for $\C$).
  \end{narenum}
\end{itemize}
\end{protocol}
}}

\vspace{8pt}

    \fbox{\parbox[t]{\adamwidth}{
        \begin{protocol}[{Multi-party Quantum Computation}]
          \label{prot:mpqc}{~}

          \begin{narenum*}
          \item \textbf{Input Phase:}
            \begin{narenum}
            \item For each $i$, player $i$ runs Top-Level Sharing with
              input $S_i$.  
            \item If $i$ is caught cheating, then some player who
              has not been caught cheating yet runs Top-Level Sharing
              (\protref{l3-vqss}), except this time with the
              one-dimensional code $\myspan{\set{\E_\C\ket 0}}$ (i.e. he
              proves that the state he is sharing is $\ket 0$).  If the
              sharing protocol fails, then another player who has not
              been caught cheating runs the protocol. There will be at
              most $t$ iterations since an honest player will always
              succeed.
            \item For each ancilla state $\ket 0$ needed for the
              circuit, some player who has not been caught cheating
              yet runs Top-Level Sharing (\protref{l3-vqss}), with the
              one-dimensional code $\myspan{\set{\E_{\C^\delta}\ket
                  0}}$\ \ or\ \ $\myspan{\set{\E_{\C^{\delta'}}\ket
                  0}}$, as needed. If the protocol fails, another
              player performs the sharing, and so forth.
            \end{narenum}
          \item \textbf{Computation Phase:} For each gate $g$ in the
            circuit, players apply the appropriate fault-tolerant
            circuit, as described in \secref{prelim}. Only the
            measurement used in Degree Reduction is not transversal.
            To measure the ancilla:
            \begin{narenum}
            \item Each player measures his component and broadcasts
              the result in the computational basis. 
            \item Let $\mathbf{w}$ be the received word. Players
              decode $\mathbf{w}$ (based on the scaled Reed-Solomon
              code $W^{\delta'}$), and obtain the measurement result
              $b$.
            \end{narenum}
          \item \textbf{Output Phase:} For the $\ith$ output wire:
            \begin{narenum}
            \item All players send their share of the output wire to
              player $i$.
            \item Player $i$ applies the decoding operator for $\C$
              and outputs the result. If decoding fails (this will
              occur only with exponentially small probability), player
              $i$ outputs $\ket 0$.               
            \end{narenum}
          \end{narenum*}
        \end{protocol}
        }}
\end{figure*}

\subsection{Distributed Computation}
\label{sec:dc}

Given the protocol of the previous section, and known fault-tolerant
techniques, there is a natural protocol for secure multi-party
computation of a circuit: have all players distribute their inputs via
the top-level sharing (\protref{l3-vqss}); apply the gates of $U$
one-by-one, using the (essentially) transversal implementation of the
gates described in \secref{prelim}; then have all players send their
share of each output to the appropriate receiver. See \protref{mpqc}.

The only sticking point in the analysis is that the fault-tolerant
procedures require some interaction when measuring a shared state.
All players measure their share and broadcast the result, applying
classical decoding to the resulting word.  If the errors occurring in
the measured ancilla were somehow correlated or entangled with errors
in the real data, one could imagine that measuring and broadcasting
them might introduce further entanglement. However, this will not be a
problem: on one hand, any errors will occur only in the cheaters'
shares, and so provide nothing beyond what the cheaters could learn
themselves; on the other hand, the honest players will discard all the
information from the broadcast except the decoded measurement result
(each honest player performs the decoding locally based on the
broadcast values, so all honest players obtain the same result).
Again, the cheaters can do this themselves.

\begin{lemma}
  Suppose that all inputs and ancillas are shared at the beginning via
  states in $\C_C$. Then the result of applying the protocol for a
  given circuit $U$, and then sending all states to an honest decoder
  $R$ is the same as sending all states to $R$ and having $R$ apply
  $U$ to the reconstructed states.
\end{lemma}

\begin{theorem}  \label{thm:mpqc-secure}
  For any circuit $U$, \protref{mpqc} is a statistically secure
   implementation of multi-party quantum computation
  as long as $t<n/6$.
\end{theorem}


\begin{proof} 
 The proof of this is by simulation, as before. The key observation
 is that when the simulator $\Sim$ is controlling the honest players,
 the adversary cannot tell the difference between the regular
 protocol and the following ideal-model simulation:
 \begin{narenum*}
 \item $\Sim$ runs the input phase as in the protocol, using $\ket 0$
   as the inputs for honest players. In this phase, if any dealer is
   caught cheating, $\Sim$ sends ``I am cheating'' to the $\ttp$ on
   behalf of that player.
    
 \item $\Sim$ ``swaps'' the cheaters' inputs with bogus data $\ket 0$,
   and sends the data to the $\ttp$. That is, he applies the
   interpolation circuit to honest players' shares to get the various
   input systems $S_i$ (for $i \in \C$), and then runs the
   interpolation circuit backwards, with the state $\ket 0$ replacing
   the original data.
    
 \item $\Sim$ now runs the computation protocol with the adversary on
   the bogus data. (Because no information is revealed on the data,
   the adversary cannot tell this from the real protocol.)
 \item $\Sim$ receives the true computation results destined to
   cheating players from $\ttp$.
 \item $\Sim$ ``swaps'' these back into the appropriate sharings, and
   sends all shares of the $\ith$ wire to player $i$ (again, he does
   this only for $i \in \C$).
 \end{narenum*}
 The proof that this simulation succeeds follows 
 from the security of the top-level sharing protocol and the previous
 discussion on fault-tolerant procedures.
\end{proof}

\section{Open Questions}
\label{sec:conc}

\an{Expendable in a pinch.}
Given our results, the most obvious open question is if \mpqc\ 
is possible when $n/6 \leq t < n/4$. 
Another natural direction of research is to find a \vqss\ protocol
with zero error. For example, the techniques of \cite{BGW88} for the
classical case do not seem to apply to the quantum setting. Finally,
one can ask what tasks are achievable when we allow cheating players
to force the abortion of the protocol (usually called an ``optimistic
protocol'').  

\section*{Acknowledgements}
Thanks to Richard Cleve for helpful discussions. A.S. thanks
Madhu Sudan for patience, support and advice.


\end{document}